\documentclass[floatfix,nofootinbib,prd,amsfonts,preprint]{revtex4-1}

\usepackage{graphicx}
\usepackage{epic}
\usepackage{eepic}
\usepackage{latexsym}
\usepackage{amssymb,amsmath}

\newcommand{\eq}[1]{(\ref{#1})}
\newcommand{\be}{\begin{equation}}
\newcommand{\ee}{\end{equation}}
\newcommand{\bea}{\begin{eqnarray}}
\newcommand{\eea}{\end{eqnarray}}

\newcommand{\vs}[1]{\vspace{#1 mm}}
\newcommand{\hs}[1]{\hspace{#1 mm}}

\newcommand{\vacr}{|0,t_0\hs{-1}>}
\newcommand{\vacl}{<\hs{-1}0,t_0|}

\def\d{\delta}

\def\e{\epsilon}

\def\fr{\frac}

\def\l{\lambda}

\def\m{\mu}

\def\S{\Sigma}

\def\th{\theta}

\def\o{\omega}

\def\del{\partial}

\let\bm=\bibitem
\def\nn{\nonumber}

\begin{document}

\title{The Functional Measure for the In-In Path Integral} 

\author{Ali Kaya}
\email[On sabbatical leave of absence from Bo\~{g}azi\c{c}i University. \\]{ali.kaya@boun.edu.tr}
\affiliation{Department of Physics, McGill University, Montreal, QC, H3A 2T8, Canada\\
and\\
Bo\~{g}azi\c{c}i University, Department of Physics, 34342, Bebek, \.{I}stanbul, Turkey\vs{15}}

\begin{abstract}

The in-in path integral of a scalar field propagating in a fixed  background is formulated in a suitable function space. The free kinetic operator, whose inverse gives the propagators of the in-in perturbation theory, becomes essentially self adjoint after imposing appropriate boundary conditions. An explicit spectral representation is given for the scalar in the flat space and the standard propagators are rederived using this representation. In this way the subtle boundary path integral over the field configurations at  the return time is handled straightforwardly. It turns out that not only the values of the forward (+) and the backward (-) evolving fields but also their time derivatives must be matched at the return time, which is mainly overlooked in the literature. This formulation also  determines  the field configurations that are included in the path integral uniquely.  We show that some of the recently suggested instanton-like solutions corresponding to the stationary phases of the cosmological in-in path integrals can be rigorously identified as limits of sequences in the function space. 

\end{abstract}

\maketitle

\section{Introduction}

The path integral formulation not only provides an equivalent alternative to the canonical quantization procedure but also it is an essential tool in understanding nonperturbative aspects of quantum theories. In addition to the vast range of applications to flat space quantum field theories, path integrals offer a  physically transparent picture of the Hawking radiation \cite{h1,h2} and moreover  it is possible to give a nonperturbative formulation of quantum gravity in terms of Euclidean path integrals \cite{h3}. 

In cosmology, however, one is not interested in transition amplitudes but rather the vacuum expectation values of  operators. In that case,  the standard formulas in perturbation theory and the usual expressions in the path integral formulation designed to yield the  transition amplitudes must be appropriately modified. This revised formalism is called in-in (or Schwinger-Keldysh or closed time) perturbation theory or correspondingly in-in path integral formalism \cite{or1,or2}. An in-in path integral giving the vacuum expectation value of an operator is different in important ways from an in-out path integral. Namely, the integration variables are doubled corresponding to forward and backward propagations in time, the boundary conditions are completely distinct and there is a subtle integration over the field configurations at the return time. In this new setup, even the equivalence of the operator and the path integral formulations in perturbation theory, which only involves Gaussian integrals,  turns out to be a nontrivial issue  \cite{w1} (see \cite{e1,e2,e3,e4,e5} for early applications of the in-in path integrals to cosmology). 

In searching for nonperturbative quantum contributions to cosmological correlations in the early Universe, the in-in path integral formulation is a valuable tool. In a recent paper \cite{a}, we apply the stationary phase approximation to the in-in path integral corresponding to a scalar field in a cosmological spacetime and find out that there exist nontrivial classical solutions contributing to the path integral, reminiscent of the instantons in Euclidean field theories. In this paper, we would like to sharpen our observations by formulating the in-in path integral as mathematically rigorous as possible.  

One way of making path integrals relatively rigorous is to make a Wick rotation the Euclidean signature. It is clear that this procedure is not applicable for general spacetimes and obviously not suitable for cosmological backgrounds where the time evolution is highly nontrivial. In that case, the ``convergence'' of the path integral can be achieved by introducing a small exponentially damping factor. This is the so called the $i\e$ prescription, which we adopt in this work. 

An in-in path integral involves three different integration variables corresponding to the field propagating forward in time (labeled by $+$), the field propagating backwards in time (labeled by $-$) and the (boundary) field, which equals  the common value of the $+$ and the $-$ fields at the return time. The integral over the boundary field has a different measure and is subtle to perform.  In the free theory, this integral can be handled by introducing a delta functional, which enforces appropriate boundary conditions for the propagators after the Gaussian path integral \cite{w1} (see also \cite{a} for an alternative treatment). Even in this case, however, one has to make sure that the operator whose inverse gives the propagators is symmetric, which requires  additional boundary conditions.  We will show that the discretized path integral, which is indeed the essential approach connecting the operator and the path integral formulations, also implies the same set of boundary conditions. This fact is mainly overlooked in the literature and might have important implications. 

One of the most straightforward ways of calculating a Gaussian path integral is to introduce the ``momentum space" mode functions for the field variables, which factorizes the functional integral into multiples of simple finite dimensional integrals. This factorization is actually achieved by the spectral representation of the kinetic operator. As noted above, the in-in kinetic operator must also be symmetric to yield the propagators of the in-in perturbation theory. This suggests to introduce a suitable {\it function space} on which the in-in kinetic operator acts, which can be elevated to a self adjoint operator. One may then introduce a suitable spectral representation to calculate the Gaussian path integral. We show that this procedure can be used to get  the in-in propagators in the flat space. In an interacting theory, once the correct propagators are obtained the connection with the standard perturbation theory in the operator formalism can easily be established. The main benefit of this new formulation in terms of a function space is that it uniquely determines  the field configurations that are included in the path integral in a nonperturbative way, which means that one may now search for other ways of approximating the full path integral around perturbatively  inaccessible configurations as in \cite{a}.
 
The plan of the paper is as follows: In the next section we consider a real scalar field propagating in a classical background and discuss the boundary conditions that make the in-in kinetic operator to be symmetric in a suitable function space. Assuming that a self adjoint extension of the operator is given, we discuss how the Gaussian path integral is performed using a spectral representation. We also show that the discretized version of the in-in path integral implies the same set of boundary conditions and thus the equivalence of the in-in path integral formula with the operator formalism only works when these are imposed.  In section \ref{sec3},  we illustrate this procedure for the flat space by calculating the propagators using the explicit spectral representation. In the same section, we also discuss the scalar propagating in a cosmological spacetime and show that the instanton-like states found  in \cite{a} can be  identified as limits of sequences in the function space. Section \ref{sec4} contains our conclusions and comments for further directions. 

\section{In-In Path integral Measure in a Function Space}

Consider a real scalar field propagating in a general curved spacetime which has the canonical action
\be\label{a}
S[\phi]=-\fr12\int d^4x\sqrt{-g}\left[\nabla_\m\phi \nabla^\m\phi+m^2\phi^2+V(\phi)\right].
\ee
Assume that the spacetime can be foliated by spacelike surfaces $\Sigma_t$ corresponding to a global time parameter $t$. It is a straightforward exercise to apply the canonical quantization procedure to \eq{a}. For the {\it free field} (or for the field  in the interaction picture) this amounts to introducing the Klein-Gordon inner product 
\be\label{kg}
(f,g)_{KG}=-i\int_\S \,d\S \,n^\m f\overleftrightarrow{\del_\mu} g^*,
\ee
where $n^\m$ is the unit normal vector to $\S$ and $d\S$ is the induced volume element. The complete set of mode functions $\m_i$ obey \cite{bd} 
\be\label{m}
\nabla^2\m_i-m^2\m_i=0,
\ee
and 
\be\label{w}
(\m_i,\m_j)_{KG}=\delta_{ij},\hs{5}(\m_i^*,\m_j^*)_{KG}=-\delta_{ij},\hs{5}(\m_i,\m_j^*)_{KG}=0.
\ee
Stoke's theorem can be used to show that the Klein-Gordon inner product is independent of the choice of $\S$. The free field operator can be expanded as
\be
\phi=\sum_i \m_ia_i+\m_i^*a_i^\dagger,
\ee
where the creation and the annihilation operators satisfy
\be
[a_i,a_j^\dagger]=\d_{ij}.
\ee
The ground state is defined by $a_i|0\hs{-1}>=0$ and the Fock space of states can be build  by acting with $a_i^\dagger$ on $|0\hs{-1}>$. In cosmology \eq{w} is equivalent to the Wronskian condition imposed on the mode functions. 

It is well known that in a general curved spacetime the vacuum state $|0\hs{-1}>$ is not uniquely defined. There are infinitely many set of mode functions obeying \eq{m} and \eq{w}. Since the mode equation \eq{m} is second order in time derivatives, $\m_i$ is uniquely determined once $\m_i(t_0)$ and $n^\m\del_\m\m_i(t_0)$ are fixed on $\S_{t_0}$ at some initial time $t_0$. Specifically when $\S=R^3$, $\m_i(t_0)$ and $n^\m\del_\m\m_i(t_0)$ can be chosen to be equal to the mode functions of the flat space, i.e $\m_k(t_0)=1/\sqrt{2(k^2+m^2)}$ and $n^\m\del_\m\m_k(t_0)=-i\sqrt{(k^2+m^2)/2}$. We denote the  vacuum state defined by suitable initial conditions at time $t_0$ as $\vacr$.

Let us consider the generating functional 
\be\label{g}
Z[J^+,J^-]=\int D\phi  \vacl  \phi,t_*\hs{-1}>_{J^-}\,<\hs{-1}\phi,t_* \vacr_{J^+},
\ee
where $\int D\phi  |\phi,t_*\hs{-1}><\hs{-1}\phi,t_*|$ is the identity operator constructed from the field variables at some {\it return time} $t_*$ and the transition amplitudes are evaluated in the presence of two independent external sources  $J^+$ and $J^-$, which are  coupled to the field variable. Differentiating \eq{g} with respect to $J^+$ and $J^-$ having different time arguments and setting $J^+=J^-=0$ give various vacuum expectation values of the field operators in the ground state $\vacr$. By introducing consecutive identity operators infinitesimally distributed in the time interval $(t_0,t_*)$ in \eq{g}, the following path integral representation can be given for the generating functional
\be\label{gb}
Z[J^+,J^-]=\int D\phi  \int \prod_{t_0}^{t_*}  {\cal D} \phi^+ {\cal D}\phi^-e^{i\left(S[\phi^+,J^+]-S[\phi^-,J^-]\right)} \Psi_0[\phi^+(t_0)]\Psi_0^*[\phi^-(t_0)] ,
\ee
where the vacuum wave functionals are defined as $\Psi_0[\phi^\pm(t_0)]=<\hs{-1}\phi^\pm(t_0)\vacr$, $D\phi$ integral is performed over field configurations at the constant time hypersurface $\Sigma_{t_*}$ and $\phi^\pm$  integrals are over all field configurations starting from time $t_0$ and ending at time $t_*$ obeying 
\be\label{bc}
\phi^+(t_*)=\phi^-(t_*)=\phi. 
\ee
It is also possible to rewrite the path integral as 
\be\label{gd}
Z[J^+,J^-]= \int  \prod_{t_0}^{t_*} {\cal D} \phi^+ {\cal D}\phi^-e^{i\left(S[\phi^+,J^+]-S[\phi^-,J^-]\right)}  \Psi_0[\phi^+(t_0)]\Psi_0^*[\phi^-(t_0)]\delta[\phi^+(t_*)-\phi^-(t_*)] , 
\ee
where now there is no restriction imposed on the integration variables, and \eq{gd}  reduces to \eq{gb} after integrating over $\phi^+(t_*)$ or $\phi^-(t_*)$.  

Let us now calculate the generating functional in the free theory and let us for the moment forget about the vacuum wave functionals and the delta functional in \eq{gd}.  This leaves a simple Gaussian path integral, which can be expressed {\it after integration by parts} as 
\be
Z_{free}[J^+,J^-]= \int \prod_{t_0}^{t_*} {\cal D} \Phi \exp\left[ \int_{t_0}^{t_*} d^4 x\,  \sqrt{-g} \left( -\fr{i}{2}\,\, \Phi^T {\bf L}\Phi+\Phi^T{\bf J}\right)\right],\label{l}
\ee
where 
\be\label{j}
{\bf L} =\left[\begin{array}{cc}L&0\\0&-L\end{array}\right],\hs{5}\Phi=\left[\begin{array}{c}\phi^+ \\\phi^- \end{array}\right], \hs{5} {\bf J}=\left[\begin{array}{c}J^+ \\ -J^- \end{array}\right]
\ee
and 
\be\label{box}
L=-\nabla_\mu\nabla^\mu+m^2.
\ee
Writing  the inverse of the operator ${\bf L}$ as
\be\label{d}
\Delta=\left[\begin{array}{cc}\Delta^{++}&\Delta^{+-}\\ \Delta^{-+}&\Delta^{--}\end{array}\right],
\ee
which obeys
\be
{\bf L}\Delta=\fr{1}{\sqrt{-g}}\left[\begin{array}{cc}\d(x,x')&0\\0&\d(x,x')\end{array}\right],
\ee
the path integral can be performed using the ``standard rules", which gives
\be\label{fr}
Z_{free}[J^+,J^-]=\exp\left(\fr{i}{2}\iint_{t_0}^{t_*}d^4x\,d^4x'\,\sqrt{-g}\,\sqrt{-g'}\,  {\bf J}^T\Delta {\bf J}\right),
\ee
where $x$ and $x'$ denote two different spacetime points. From the very definition of the generating functional, the propagators can be determined in terms of the free field vacuum expectation values as 
\bea
&&\Delta^{++}(x,x')=i\vacl T\phi(x)\phi(x')\vacr, \nn\\
&&\Delta^{--}(x,x')=i\vacl \overline{T}\phi(x)\phi(x')\vacr, \label{prop}\\
&&\Delta^{-+}(x,x')=i\vacl \phi(x)\phi(x')\vacr,\nn\\
&&\Delta^{+-}(x,x')=i\vacl \phi(x')\phi(x)\vacr,\nn
\eea
where $T$ and $\overline{T}$ denote time and anti-time orderings, respectively. Note that the Green functions  obey
\bea
&&\Delta^{+-}(x,x')=\Delta^{-+}(x',x),\nn\\
&&\Delta^{++}(x,x')=\Delta^{++}(x',x),\label{sym}\\
&&\Delta^{--}(x,x')=\Delta^{--}(x',x),\nn
\eea
which shows that $\Delta$ is a {\it symmetric kernel.} 

Even though the above derivation looks trivial, there are important subtleties that should be addressed. The existence of the off-diagonal terms in \eq{d} begs for an explanation since one would naively write a diagonal Green function corresponding to a set of decoupled path integrals. Fortunately, this point is clarified in the Appendix of \cite{w1}, which explains that the form of \eq{d} is fixed by the delta functional in \eq{gd} that couples $\phi^+$ and $\phi^-$ integrals. Indeed, as shown in \cite{w1} the sole effect of the delta functional is to impose 
\bea
\Delta^{++}(t_*)=\Delta^{-+}(t_*),\nn\\
\Delta^{--}(t_*)=\Delta^{+-}(t_*).
\eea
On the other hand, at least in $t_0\to-\infty$ limit the vacuum wave functionals are supposed to yield a suitable $i\e$ prescription such that $\Delta^{++}$ and $\Delta^{--}$ propagators are given in terms of the time and anti-time ordered products of the free fields as in \eq{prop} (see e.g. \cite{wb}). 

However, additional conditions are still needed to make the Gaussian path integral well defined and to determine the Green functions uniquely. In obtaining the second order kinetic operator in \eq{l}, both actions $S[\phi^\pm]$ are integrated by parts and the surface terms are discarded. Although in many applications $t_0\to-\infty$ limit is taken and the surface terms vanish after imposing suitable fall of conditions at the asymptotic past infinity, there is no  reason for the surface terms to vanish at $t_*$. The absence of the surface terms is crucial to have a {\it symmetric differential operator} so that the Gaussian path integral becomes well defined and the corresponding Green function becomes symmetric in its arguments as in \eq{sym}.  Although in a finite dimensional Gaussian integral the quadratic integrand picks up the symmetric part of the matrix automatically, this is no longer true for the infinite dimensional case. 
 
Since the original path integral derivation of the generating functional requires \eq{bc} to hold, the surface terms at $t_*$ vanish if and only if the following boundary conditions are imposed 
\bea\label{ybc1}
&&\phi^+(t_*)=\phi^-(t_*), \\
&&n^\mu\del_\mu \phi^+(t_*)=n^\mu\del_\mu\phi^-(t_*). \label{ybc2}
\eea
The free Gaussian path integral becomes well defined after additional boundary conditions  are implemented at $t_0$. Although the first condition \eq{ybc1} is already encountered in the path integral derivation, the second condition \eq{ybc2} might appear surprising and can be thought to ruin the equivalence of the path integral and the operator formalisms. As we will discuss shortly,  the opposite turns out to be true, namely \eq{ybc2} is enforced by the equivalence. As noted above, \eq{ybc2} is a hidden assumption that is used to obtain the propagators. 

Despite the fact that the Gaussian path integral is uniquely determined after imposing \eq{ybc1} and \eq{ybc2}, one may still search for a more direct way of calculating the integral. Let us recall, for example, how the propagator of the harmonic oscillator is obtained in the in-out formalism. The path integral representation of the generating functional is given by  
\be\label{hj}
z[j]=\int Dq\exp\left(\int_{-\infty}^{\infty}dt [\fr{i}{2}\dot{q}^2-\fr{im^2}{2}q^2+jq]\right).
\ee
To perform the Gaussian integral one can Fourier transform the variables as 
\bea
q(t)=\fr{1}{\sqrt{2\pi}}\int_{-\infty}^{\infty}dE\,e^{iEt}\, \tilde{q}(E),\\
j(t)=\fr{1}{\sqrt{2\pi}}\int_{-\infty}^{\infty}dE\,e^{iEt}\, \tilde{j}(E).
\eea
In the transformed variables, the integral becomes 
\be
z[j]=\int \prod_E D\tilde{q}(E)\exp\left(\int_{-\infty}^{\infty}\left[\fr{i}{2}\tilde{q}(E)[E^2-m^2]\tilde{q}(-E)+\tilde{j}(-E)q(E)\right]dE\right).
\ee
Therefore, the whole path integral factorizes into multiples of one dimensional Gaussian integrals for each $E$. While this  one dimensional integral over $E$ does not converge,  it can be shown that in and out vacuum wave functionals, which are neglected in \eq{hj}, give rise to $i\e$ terms guaranteeing the convergence of the $E$-integral. Introducing these $i\e$ terms, the final result can be written as (see e.g. \cite{srd})
\be\label{je}
z[j]=\exp\left(\fr{i}{2}\int_{-\infty}^{\infty}dE\fr{\tilde{j}(E)\tilde{j}(-E)}{E^2-m^2+i\e}\right).
\ee
It is now a straightforward exercise to express the propagator in the coordinate basis, namely in the $t$-variable, by applying the inverse Fourier transformation to \eq{je}, which should give the standard result. 

Naturally, a similar procedure is expected to be applicable to calculate the in-in path integral. The harmonic oscillator example shows how this can be achieved. Mathematically speaking, the Fourier transformation, or the set of the functions $e^{iEt}/\sqrt{2\pi}$, give an explicit spectral representation of the kinetic operator $d^2/dt^2$ in \eq{hj}. There is a continuos  spectrum\footnote{Actually the spectrum is given by $\l=-E^2$ and for each eigenvalue $\l$ there correspond to two eigenvectors.} $E\in(-\infty,\infty)$ and the paths included in the path integral belong to the space of square integrable functions in the real line: $q(t)\in L^2(-\infty,\infty)$. This guarantees that no surface terms appear after integrations by parts in \eq{hj} and the operator $d^2/dt^2$ becomes symmetric, which has a unique self adjoint extension in $L^2(-\infty,\infty)$. 

Analogously, to evaluate the free in-in path integral \eq{l},  one can introduce the Hilbert space of {\it doublets} of square integrable real\footnote{It is straightforward to construct the Hilbert space for complex functions. Since we are considering a real scalar field, we prefer to work with a real Hilbert space. To clarify the notation, $(2,2)$ stands for square integrable doublets of functions.} functions,  $L^{(2,2)}(t_0,t_*)$, with the following inner product
\be\label{ip}
<\hs{-1}\Phi_1|\Phi_2\hs{-1}>=\int_{R} d^4 x\, \sqrt{-g}\, \Phi_1^T \Phi_2,
\ee
where  $R$ is the region bounded by the times $t_0$ and $t_*$, i.e. $R=\Sigma\times(t_0,t_*)$, $\Phi_1=(\phi_1^+,\phi_1^-)$ andÊ $\Phi_1=(\phi_2^+,\phi_2^-)$. The kinetic operator 
\be\label{j2}
{\bf L} =\left[\begin{array}{cc}L&0\\0&-L\end{array}\right],
\ee
where $L=-\nabla_\mu\nabla^\mu+m^2$, is a second order differential operator acting on $L^{(2,2)}(t_0,t_*)$. To avoid complications related to the vacuum wave functionals, let us take  $t_0=-\infty$. In that case, ${\bf L}$ becomes a {\it symmetric operator} provided \eq{ybc1} and \eq{ybc2} are imposed as boundary conditions. Assume now that these boundary conditions ensure ${\bf L}$ to be essentially self adjoint operator in 
$L^{(2,2)}(-\infty,t_*)$. In that case, one can introduce a complete set of real (generalized) eigenfunctions 
\be
u(\l)=\left[\begin{array}{c}u^+(\l,x)\\ u^-(\l,x)\end{array}\right]
\ee
satisfying  
\be
{\bf L} u(\l)=\l\, u(\l)
\ee
and the boundary conditions  \eq{ybc1} and \eq{ybc2} 
\bea\label{ub1}
&&u^+(t_*)=u^-(t_*), \\
&&n^\mu\del_\mu u^+(t_*)=n^\mu\del_\mu u^-(t_*). \label{ub2}
\eea
The eigenfunctions $u(\l)$ span  $L^{(2,2)}(-\infty,t_*)$ and any $\Phi\in L^{(2,2)}(-\infty,t_*)$ obeying \eq{ybc1} and \eq{ybc2} can be expanded as  
\be
\Phi=\int d\m(\l)\, \tilde{\phi}(\l)\,u(\l),
\ee 
where $d\mu(\l)$ denotes a measure in the set of eigenvalues. The eigenfunctions can be normalized as
\be
<\hs{-1}u(\l)|u(\l')\hs{-1}>=\d(\l-\l')
\ee
so that the inner product becomes
\be
<\hs{-1}\Phi|\Phi\hs{-1}>=\int d\mu(\l)\,\tilde{\phi}(\l)^2.
\ee
The eigenfunctions must also obey the completeness relation
\be\label{c1}
\int d\mu(\l)\, u(\l,x)u^T(\l,x') =\fr{1}{\sqrt{-g}}\d(x,x').
\ee
Note that while $\Phi$ denotes a doublet of functions, $\tilde{\phi}(\l)$ is a single real function. 

Using the spectral representation, the generating functional can be expressed as 
\bea
Z_{free}[J^+,J^-]&=& \int \prod_{-\infty}^{t_*} {\cal D} \Phi \exp\left(-\fr{i}{2}\,\, <\hs{-1}\Phi|{\bf L}|\Phi\hs{-1}>+<\hs{-1}\Phi|{\bf J}\hs{-1}>\right),\nn\\
&=&\int D\tilde{\phi}(\l) \exp\left(\int d\mu(\l)\left[-\fr{i}{2}\,\l\tilde{\phi}(\l)^2+\tilde{\phi}(\l)\tilde{J}(\l)\right]\right), \label{38}
\eea
where $\tilde{J}(\l)=<\hs{-1}u(\l)|{\bf J}>$. Since the spectral functions $u(\l)$ obey \eq{ub1},  the delta functional in \eq{gd} is automatically satisfied (actually the path integral is over a more restricted set of of functions, which also obey \eq{ybc2}). 

As in the case of harmonic oscillator, the Gaussian path integral factorizes into one dimensional ordinary integrals, whose convergence can be fulfilled by introducing a small imaginary piece $\l\to\l-i\e$. Evaluating the integral over $\tilde{\phi}(\l)$ one finds
\be\label{zs}
Z_{free}[J^+,J^-]= \exp\left(-\fr{i}{2}\int d\mu(\l)\left[\fr{\tilde{J}(\l)^2}{\l-i\e}\right]\right).
\ee
By comparing \eq{zs} to \eq{fr} and using the completeness relation \eq{c1} the propagator can be expressed as   
\be\label{gf}
\Delta(x,x')=-\int d\mu(\l)\, \fr{u(\l,x)u^T(\l,x')}{\l-i\e}.
\ee
The vacuum wave functionals entering the path integral at  $t_0=-\infty$ as in \eq{gb} are supposed to yield the $i\e$ term in the path integral. One therefore sees that the free in-in path integral can be naturally  calculated using the spectral representation of ${\bf L}$. Unless $u^+(\l)=0$ or $u^-(\l)=0$ for all $\l$, there appears off-diagonal terms in \eq{gf}. 

The inner product \eq{ip} should not be confused with the Klein-Gordan inner product \eq{kg}. Similarly, the boundary conditions \eq{ybc1} and \eq{ybc2}, which are imposed for the Gaussian path integral and depend on the arbitrary return time $t_*$, are not directly related to the boundary conditions obeyed by the canonical fields in the quantization procedure. Both the inner product \eq{ip} and the conditions \eq{ybc1} and \eq{ybc2} are implemented to have a mathematically well defined Gaussian path integral. 

Till now, we only consider the free theory and discuss how one can obtain the in-in propagators in an alternative but mathematically more rigorous way. In the interacting theory, in case when perturbation theory is applicable, the set of fields that are integrated out should still be same so that the Gaussian integral yields the correct in-in propagators consistent with the operator formalism. When the theory becomes strongly coupled, even though the perturbation theory essentially breaks down, one can still imagine that the exact results must match  the sum of the infinite perturbative series. In other words, one would expect the set of fields that are integrated out to be independent of whether the theory is weakly or strongly coupled (of course the value of the path integral depends on the coupling constants of the theory and weakly and strongly coupled limits are expected to give very different results.) Therefore, the {\it exact} path integral for the generating functional 
\be\label{gbb}
Z[J^+,J^-]= \int \prod_{-\infty}^{t_*}  {\cal D} \phi^+ {\cal D}\phi^-e^{i\left(S[\phi^+,J^+]-S[\phi^-,J^-]\right)} \Psi_0[\phi^+(-\infty)]\Psi_0^*[\phi^-(-\infty)] ,
\ee
must be a sum over all in-in paths that belong to $L^{(2,2)}(-\infty,t_*)$ obeying the boundary conditions \eq{ybc1} and \eq{ybc2}: 
\be \label{cr}
(\phi^+,\phi^-)\in L^{(2,2)}(-\infty,t_*), \hs{5}\phi^+(t_*)=\phi^-(t_*), \hs{5}n^\mu\del_\mu \phi^+(t_*)=n^\mu\del_\mu\phi^-(t_*). 
\ee
Although the path integral measure is not defined rigorously, \eq{cr} gives a precise nonperturbative criteria about the configurations included in the sum. As we will see in the next section, this is important in applying approximation methods in evaluating the in-in path integrals.  

Our discussion so far is based on the free path integral and in the last paragraph we made a generalization to the exact theory. The main strategy was to be consistent with the perturbation theory or more precisely with the operator formalism. As is well known, the first boundary condition in \eq{cr}, i.e. $\phi^+(t_*)=\phi^-(t_*)$, arises naturally when one derives the in-in path integral from the operator formalism (this can be directly seen from \eq{g}). The second condition in \eq{cr}, on the other hand,  is merely deduced from the  consistency of the free Gaussian path integral and we argue that it should still hold in the interacting theory. It is thus crucial  to see how this condition arises in the full path integral, or if not how the equivalence of the path integral and the operator formalisms still holds. To analyze the issue more carefully, let us consider the discretized path integral for \eq{gbb}, which is indeed the main object connecting the operator and the path integral formalisms. In that case, assuming that the metric is of the form $ds^2=-dt^2+d\S^2$ and the scalar has the standard action \eq{a}, \eq{gbb} can be written as
\be\label{dc}
 Z=\int d\phi_{t_*}  d\phi^+_{t_*-\e} d\phi^-_{t_*-\e}...\exp\left[i\e \sum  \left(\fr{\phi_{t_*}-\phi^+_{t_*-\e}}{\e}\right)^2 - \left(\fr{\phi_{t_*}-\phi^-_{t_*-\e}}{\e}\right)^2+...  \right],
\ee
where $\e$ is a discretization parameter (not to be confused with $\e$ in the $i\e$ prescription) and we  suppress all but the time coordinate as the index. In \eq{dc} we only write the terms that depend on the boundary variable $\phi_{t_*}$, which are precisely  the terms containing  a time derivative in the action.  Note that the terms involving the spatial gradients and the potential cancel out each other between $\pm$ branches at time $t_*$. On the other hand, the source couplings turn out to be $\e$-suppressed compared to \eq{43} or one may take $t_*$ large enough so that the sources vanish, i.e. $J^\pm(t)=0$ for $t\geq t_*$, as we assume in \eq{dc}. We thus see from \eq{dc} that the boundary integral has the form
\be\label{43}
Z=\int d\phi_{t_*} ...\exp\left[2i  \phi_{t_*}  \left(\fr{\phi^-_{t_*-\e}-\phi^+_{t_*-\e}}{\e}\right)+...  \right],
\ee
which implies 
\be
\lim_{\e\to 0}\fr{\phi^-_{t_*-\e}-\phi^+_{t_*-\e}}{\e}=\lim_{\e\to 0}\fr{(\phi_{t_*}-\phi^+_{t_*-\e})}{\e}-\fr{(\phi_{t_*}-\phi^-_{t_*-\e})}{\e}=\dot{\phi}^+(t_*)-\dot{\phi}^-(t_*)=0. 
\ee
This is nothing but \eq{ybc2} written in coordinates normal to the surface $\S$. It is  remarkable that the path integral imposes the same conditions that are necessary to make the kinetic operator symmetric. 

Summarizing our results, we find that to calculate the in-in generating functional one must integrate over all field configurations in $L^{(2,2)}(-\infty,t_*)$ obeying \eq{ybc1} and \eq{ybc2}. In perturbation theory, the same conditions are required for the kinetic operator to be symmetric and assuming there exist a self adjoint extension, the propagators can be calculated using the spectral functions. Recall that the first condition arises due to the identity operator in \eq{g} and the second follows from the boundary integral as shown above.  The condition \eq{ybc2}  is mainly overlooked in the literature (see e.g. \cite{yeni} for an exception)  and it should clearly be considered in determining the effects of boundary terms to the cosmological correlation functions (see \cite{b1,b2,b3,b4,b5}). 

\section{The flat space and cosmological examples}  \label{sec3}

In the previous section, we discuss how the free in-in Gaussian path integral can be performed using the spectral representation of the in-in kinetic operator. In this section, we would like to carry out an explicit  computation for a scalar propagating in the flat space with the Lorentzian metric
\be
ds^2=-dt^2+dx^2+dy^2+dz^2.
\ee
Without loss of any generality we take $t_*=0$ and consider $t_0=-\infty$ case. As usual, by applying a Fourier transformation in $R^3$ with the  functions $\exp(i{\bf x}.{\bf k})/(2\pi)^{3/2}$, the in-in kinetic operator can be transformed into 
\be\label{jj}
{\bf L} =\left[\begin{array}{cc}\fr{d^2}{dt^2}+\o^2&0\\0&-(\fr{d^2}{dt^2}+\o^2)\end{array}\right],
\ee
where $\o^2=k^2+m^2$. To our knowledge the spectral analysis of this operator in the half line $(-\infty,0)$ is not studied before. 

Since the spatial dependence of the functions is handled by the Fourier transformation, it is enough to consider a reduced Hilbert space, one for each wave number ${\bf k}$, consisting of doublets of time dependent functions $\Phi(t)=(\phi^+(t),\phi^-(t))$ with the inner product
\be\label{rip}
<\hs{-1}\Phi_1|\Phi_2\hs{-1}>=\int_{-\infty}^0 dt \, \Phi_1^T(t) \Phi_2(t).
\ee
The boundary conditions of interest that make ${\bf L}$ symmetric are given by 
\bea\label{rbc1}
&&\phi^+(0)=\phi^-(0), \\
&&\dot{\phi}^+(0)=\dot{\phi}^-(0). \label{rbc2}
\eea
We would like to see whether  {\bf L} is essentially self adjoint or not. Since ${\bf L}$ is real the upper and lower deficiency indices equal and it suffices to look for normalizable solutions of 
\be
{\bf L} \Phi(\l)=i\, \Phi(\l),
\ee
which can be solved as
\bea
&&\phi^+=a_1e^{z_1t}+a_2e^{-z_1t}\\
&&\phi^-=a_3 e^{z_2t}+a_4e^{-z_2t},
\eea
where $z_1^2=i-\o^2$ and $z_2^2=-i-\o^2$. Normalizability of the solutions requires $a_2=a_3=0$ (note that $t<0$) and the  boundary conditions \eq{rbc1} and \eq{rbc2} give $a_1=a_4=0$, which shows that there are no eigenfunctions with imaginary  eigenvalues and thus ${\bf L}$ is essentially self adjoint.   

To determine the generalized eigenfunctions of ${\bf L}$,  one needs to solve
\be
{\bf L} u(\l)=\l\, u(\l).
\ee
Writing $u(\l)=(u^+,u^-)$, this implies 
\bea
&&\fr{d^2 u^+}{dt^2}=(\l-\o^2)u^+,\label{u1}\\
&&\fr{d^2 u^-}{dt^2}=-(\l+\o^2)u^-.\label{u2}
\eea
The solutions are given either by  $\sin$ and $\cos$ functions or by the exponential. Although the oscillating functions are allowed, which would yield delta function normalization of the eigenfunctions, the exponentially growing pieces should be discarded. The solutions for different values of $\l$ can be found as
\bea
&&\l>\o^2:\hs{21} u(\l)=\left[\begin{array}{l} b_1\exp\left(t{\sqrt{\l-\o^2}}\right) \\
b_2\sin\left(t\sqrt{\l+\o^2}\right)+b_3\cos\left(t\sqrt{\l+\o^2}\right)\end{array}\right],\nn\\
&&\o^2\geq\l\geq-\o^2:\hs{8}
u(\l)=\left[\begin{array}{l} c_1\sin\left(t\sqrt{-\l+\o^2}\right)+c_2\cos\left(t\sqrt{-\l+\o^2}\right)\\
 c_3\sin\left(t\sqrt{\l+\o^2}\right)+c_4\cos\left(t\sqrt{\l+\o^2}\right) 
 \end{array}\right],\label{sol}\\
&&\l<-\o^2:\hs{18}
u(\l)=\left[\begin{array}{l} d_1\sin\left(t\sqrt{-\l+\o^2}\right)+d_2\cos\left(t\sqrt{-\l+\o^2}\right)\\
 d_3\exp\left(t{\sqrt{-\l-\o^2}}\right) \end{array}\right].\nn
\eea
The unknown coefficients must be fixed by the boundary conditions \eq{rbc1} and \eq{rbc2} together with the  orthonormality and completeness of the eigenfunctions. As we present the important  details in the Appendix, these requirements determine the coefficients uniquely. For $\l>\o^2$ and $\l<-\o^2$, there is a single eigenfunction for each $\l$: 
\bea
&&\l>\o^2:\hs{17} u(\l)=\fr{(\l+\o^2)^{1/4}}{\sqrt{2\pi\l}}}{\left[\begin{array}{l} \exp\left(t{\sqrt{\l-\o^2}}\right) \\
\sqrt{\fr{\l-\o^2}{\l+\o^2}}\sin\left(t\sqrt{\l+\o^2}\right)+\cos\left(t\sqrt{\l+\o^2}\right)\end{array}\right],\label{f1}\\
&&\l<-\o^2:\hs{14} u(\l)=\fr{(-\l+\o^2)^{1/4}}{\sqrt{-2\pi\l}}}{\left[\begin{array}{l} \sqrt{\fr{-\l-\o^2}{-\l+\o^2}}\sin\left(t\sqrt{-\l+\o^2}\right)+\cos\left(t\sqrt{-\l+\o^2}\right) \\ 
\exp\left(t{\sqrt{-\l-\o^2}}\right)
\end{array}\right] .\nn
\eea
On the other hand, for $-\o^2\leq \l\leq \o^2$ there are two eigenfunctions for each $\l$, which are given by 
\bea
&&\o^2\geq\l\geq-\o^2:\hs{5}
u_1(\l)=\left(\fr{\sqrt{\l+\o^2}-\sqrt{-\l+\o^2}}{2\pi\l}\right)^{1/2}\left[\begin{array}{l}\left(\fr{\l+\o^2}{-\l+\o^2}\right)^{1/4} \sin\left(t\sqrt{-\l+\o^2}\right)\\
\left(\fr{-\l+\o^2}{\l+\o^2}\right)^{1/4}\sin\left(t\sqrt{\l+\o^2}\right)
 \end{array}\right],\nn \\
&&\o^2\geq\l\geq-\o^2:\hs{5}
u_2(\l)=\left(\fr{\sqrt{\l+\o^2}-\sqrt{-\l+\o^2}}{2\pi\l}\right)^{1/2}\left[\begin{array}{l}\ \cos\left(t\sqrt{-\l+\o^2}\right)\\
\cos\left(t\sqrt{\l+\o^2}\right)
 \end{array}\right].\label{f2}
 \eea
As shown in the Appendix, the eigenfunctions are properly normalized with respect to the norm \eq{rip} 
\be
<u(\l)|u(\l')>=\d(\l-\l')\label{norm}
\ee
and they are also complete
\be
\int_{-\infty}^{\infty}\,d\l\, u(\l,t)u^T(\l,t')=\left[\begin{array}{cc}\d(t-t')&0\\0&\d(t-t')\end{array}\right],\label{comp}
\ee
where in the range $-\o^2\leq \l\leq \o^2$ both of the eigenfunctions $u_1(\l)$ and $u_2(\l)$ in  \eq{f2} must be included in the integral.  

Once the complete set of eigenfunctions are obtained, the Green function can be calculated using \eq{gf} as
\be\label{gu}
\Delta(t,t')=-\int_{-\infty}^{\infty} d\l\, \fr{u(\l,t)u^T(\l,t')}{\l-i\e}.
\ee
Similar to the harmonic oscillator example, this computation can be carried out using the contour integration techniques  (see Appendix), which gives the in-in propagator in momentum space
\be\label{fp}
\Delta(t,t')=-i\left[\begin{array}{cc}  \fr{\th(t-t')}{2\o}e^{i\o(t-t')}+ \fr{\th(t'-t)}{2\o}e^{-i\o(t-t')} & \fr{1}{2\o}e^{i\o(t-t')}\\
 \fr{1}{2\o}e^{-i\o(t-t')} & \fr{\th(t'-t)}{2\o}e^{i\o(t-t')}+ \fr{\th(t-t')}{2\o}e^{-i\o(t-t')} 
\end{array}\right].
\ee
This shows that the procedure outlined in the previous section for the calculation of the Gaussian in-in path integrals is consistent with the operator formalism. In particular, if the second condition \eq{rbc2} would not be imposed then \eq{fp} could not  be obtained. 

In principle, the above discussion can easily be generalized to cosmological spacetimes with the metric
\be
ds^2=-dt^2+a(t)^2(dx^2+dy^2+dz^2).
\ee
After applying the Fourier transformation along the spatial directions, the in-in kinetic operator becomes
\be\label{lc}
{\bf L} =\left[\begin{array}{cc}\fr{d^2}{dt^2}+\fr{3\dot{a}}{a}\fr{d}{dt}+\o(t)^2&0\\0&-\left(\fr{d^2}{dt^2}+\fr{3\dot{a}}{a}\fr{d}{dt}+\o(t)^2\right)\end{array}\right],
\ee
where $\o(t)^2=k^2/a^2+m^2$. As before, one may define a reduced function space on which ${\bf L}$ acts  consisting of doublets of time dependent functions $\Phi(t)=(\phi^+(t),\phi^-(t))$ with the inner product
\be\label{ripc}
<\hs{-1}\Phi_1|\Phi_2\hs{-1}>=\int_{-\infty}^0 dt \,a(t)^3 \, \Phi_1^T(t) \Phi_2(t).
\ee
Note that $a^3$ term in \eq{ripc} plays the role of a weight function. 

It is easy to see that if appropriate fall off conditions are imposed at the asymptotic past infinity, ${\bf L}$ in \eq{lc} becomes a symmetric operator provided the usual boundary conditions \eq{rbc1} and \eq{rbc2} are imposed at $t=t_*=0$. Since the mode functions are not known for general $a(t)$, it is not possible to prove that \eq{lc} is essentially self adjoint by looking at the deficiency indices. On the other hand, for de Sitter space in the Poincare patch that has $a=\exp(Ht)$, the upper deficiency indices are determined by the normalizable solutions of 
\bea
&&\fr{d^2\phi^+}{dt^2}+3H\fr{d\phi^+}{dt}+\left(e^{-2Ht}k^2+m^2\right)\phi^+=i\phi^+,\\
&&\fr{d^2\phi^-}{dt^2}+3H\fr{d\phi^-}{dt}+\left(e^{-2Ht}k^2+m^2\right)\phi^-=-i\phi^-.
\eea
These can be solved as
\bea
&&\phi^+=e^{-3Ht/2}\left[c_1 H_\mu^{(1)}(k\eta)+c_2H_\mu^{(2)}(k\eta)\right],\\
&&\phi^-=e^{-3Ht/2}\left[c_3 H_\nu^{(1)}(k\eta)+c_2H_\nu^{(2)}(k\eta)\right],
\eea
where $H^{(1)}$ and $H^{(2)}$ denote the Hankel functions of the first and the second kinds, the conformal time $\eta$ is given by $\eta=\exp(-Ht)/H$, $\mu=\sqrt{9/4-m^2/H^2+i/H^2}$ and $\nu=\sqrt{9/4-m^2/H^2-i/H^2}$. Using the asymptotic form of the Hankel functions it is easy to see that the solutions are not normalizable and thus the symmetric operator \eq{lc} is indeed essentially self adjoint in de Sitter space. 

Till now, we have basically studied the spectral properties of the in-in kinetic operator, which is crucial in perturbation theory. However, as discussed at the end of the previous section, the formulation of the in-in path integral in a suitable function space does not rely on perturbation theory. We would like to illustrate 
how our construction can be used to infer nonperturbative properties, more specifically we would like to consider the stationary phase approximation and show how some of the instanton-like solutions constructed in \cite{a} can be rigorously  understood. For that,  take a self interacting real scalar  field with the potential (see Fig. \ref{fg1}) 
\be\label{pot}
V=\fr{\l}{4}\phi^2(\phi-\phi_0)^2
\ee
in the Poincare patch of the de Sitter space, i.e. $a=\exp(Ht)$. We focus on the {\it zero mode} in the path integral, which is a sum over all square integrable doublet of functions $\Phi$ with the norm
\be\label{pc}
<\hs{-1}\Phi|\Phi\hs{-1}>=\int_{-\infty}^0 dt \,e^{3Ht} \, \Phi^T(t) \Phi(t).
\ee
We would like to show that if $\phi_{cl}$ is an arbitrary solution of the classical equations of motion 
\be\label{em}
\ddot{\phi}_{cl} +3H\dot{\phi}_{cl}+\fr{\del V}{\del \phi_{cl}}=0,
\ee
then the in-in field 
\be\label{iif}
\Phi=\left[\begin{array}{c}\phi_{cl}\\ \phi_{cl}\end{array}\right]
\ee
is included in the path integral for the generating functional. This amounts to prove that  the following integral  converges for arbitrary initial data:
\be\label{conv}
\int_{-\infty}^0 dt \,e^{3Ht} \, \phi_{cl}^2.
\ee
It is known that \eq{em} gives oscillations that are damped by the expansion of the Universe. Viewing evolution backwards in time, $\phi_{cl}$ tends to grow as $t\to -\infty$, which would fight for the exponential factor in \eq{conv}. Let us, for the moment, forget about the mass term and the expansion of the Universe. Then \eq{em} becomes
\be
\ddot{\phi}_{cl} +\l \phi_{cl}^4=0,
\ee
which can be solved as 
\be\label{sn}
\phi_{cl}=A\,{\textrm Sn} \left(A \sqrt{\fr{\l}{2}},-1\right),
\ee
where Sn is the Jacobi elliptic function, which periodically oscillates between $-1$ and $1$. To determine the solution with the Hubble term we note that the amplitude $A$ also determines the frequency of the oscillations in \eq{sn}. Since we are interested in the asymptotic behavior of the fields as $t\to-\infty$ corresponding to large field fluctuations, i.e. $\phi_{cl}\sim A\gg H$, the frequency of the oscillations becomes much larger than the expansion rate in that limit. Therefore, one may try an approximate solution with slowly varying amplitude $A(t)$. Using such an ansatz in the equations of motion
\be\label{sne}
\ddot{\phi}_{cl} +3H\dot{\phi}+\l \phi_{cl}^4=0,
\ee
one can find the following asymptotic solution
\be\label{af}
\phi_{cl}\simeq A_0\exp(-Ht)\,{\textrm Sn}\,\left(\fr{\exp(-Ht)}{H}\sqrt{\fr{\l}{4}},-1\right),\hs{5} t\to-\infty.
\ee
Including the mass term to \eq{sne} does not alter this asymptotic form since the mass term becomes eventually negligible compared to $\phi^4$ term as $t\to-\infty$. Eq. \eq{af} shows that the integral \eq{conv} converges and thus all classical solutions are included in the in-in path integral in the form \eq{iif}. 

\begin{figure}
\centerline{\includegraphics[width=8cm]{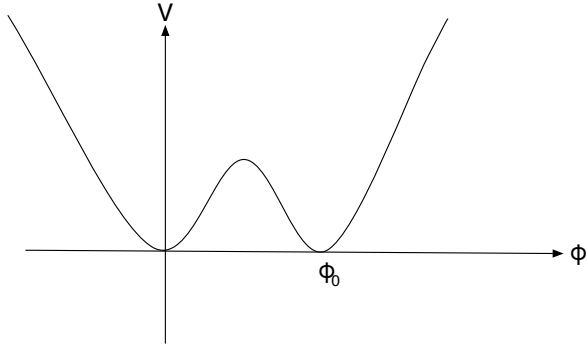}}
\caption{The scalar potential $V=\fr{\l}{4}\phi^2(\phi-\phi_0)^2$.} 
\label{fg1}
\end{figure}

In \cite{a}, the stationary phase approximation is applied to calculate the in-in path integral for the generating functional and it is shown that the stationary phases are given by \eq{iif} where  $\phi_{cl}(-\infty)$ equals the expectation value of the field operator in the vacuum state  at $t=-\infty$. For the potential \eq{pot}, there are two possible vacuum states. Assuming that one expands the theory around the vacuum $\phi=0$, the stationary phases must obey
\be\label{sf}
\phi(-\infty)=0.
\ee
As discussed in \cite{a}, the classical solutions with the initial data 
\be\label{inst}
\phi(-\infty)=0,\hs{5}\dot{\phi}(-\infty)=v_0
\ee
can interpolate between the vacua $\phi=0$ and $\phi=\phi_0$. Namely, even one is expanding around the vacuum at $\phi=0$, the vacuum at $\phi=\phi_0$ also contributes to the generating functional via these classical configurations. They are very similar to the instantons of Euclidean field theories. 

Although the construction of instanton-like solutions presented in \cite{a}  is physically viable, which can be attributed to the spontaneous fluctuations around the vacuum, giving  initial data at $t=-\infty$ can be mathematically doubtful. Let us therefore analyze the problem in a different way. Fix an arbitrary (negative) time $\tilde{t}$  and consider the classical solution with the initial data
\be
\phi_1(\tilde{t})=0,\hs{5}\dot{\phi}_1(\tilde{t})=v_0,
\ee
which we identify as the first member of a series of solutions. We define the $(n+1)$'th solution by  shifting the time argument of the $n$'th solution by $1/n$: 
\be
\phi_{n+1}(t)=\phi_n\left(t-\fr{1}{n}\right).
\ee
Since the equations of motion \eq{em} does not depend on $t$ explicitly, $\phi_{n+1}$ is a solution provided $\phi_n$ is a solution. Since $\phi_1$ is a solution by construction, all functions in the sequence must solve equations of motion. As proved above, all classical solutions are square integrable and thus belong to the Hilbert space in which the path integral is calculated. 

It is clear that as $n$ gets larger $\phi_n$ approaches the instanton-like configuration \eq{inst}. Moreover, 
\bea
\lim_{n\to\infty}\left|\phi_{n+1}-\phi_n\right|&=&\lim_{n\to\infty}\int_{-\infty}^0dt\,e^{3Ht}\left[\phi_{n+1}(t)-\phi_n(t)\right]^2\nn\\
&=&\lim_{n\to\infty}\int_{-\infty}^0dt\,e^{3Ht}\left[\phi_{n}(t-\fr{1}{n})-\phi_n(t)\right]^2=0,
\eea
which shows that the sequence has an accumulation point. By completeness, the accumulation point must also be included in the Hilbert space, which shows that the in-in path integral can be expanded around it. Depending on the value of the initial velocity $v_0$, one has $\lim_{n\to\infty}\phi_n=0$ or $\lim_{n\to\infty}\phi_n=\phi_0$, which are identified in \cite{a} as the instanton-like configurations for the potential \eq{pot}. 
Namely, we see that even though for finite $n$, $\phi_n$  does not obey \eq{sf} and thus  the field $\Phi=(\phi_n,\phi_n)$ is not a suitable point for stationary phase approximation around the vacuum at $\phi=0$, the approximation gets better and better as $n$ gets larger. Alternatively, in spite of the fact that $\phi=\phi_0$ does not obey the boundary condition \eq{sf}, it is possible to find a sequence of solutions in the Hilbert space, whose members satisfy \eq{sf} and approach arbitrarily close to $\phi=\phi_0$ solution in the limit. 

\section{Conclusions} \label{sec4} 

In this paper, we reformulate the in-in path integral corresponding to a real scalar field by defining a suitable function space. This construction allows the calculation of  the free Gaussian path integral by using the spectral representation of the in-in kinetic operator. Since the  in-in states that are integrated out belong to a specified function space, one has a well defined criteria if a given configuration can be used for the expansion of the path integral around it. As discussed above, this is important in applying the stationary phase approximation to the in-in path integrals. 

An important point emphasized in this work is that there are two boundary conditions that must be imposed to make the in-in kinetic operator symmetric. The first condition \eq{ybc1} is obvious in the path integral derivation of the generating functional. After imposing \eq{ybc1}, the second condition \eq{ybc2} is also required to have a symmetric in-in kinetic operator, which is certainly necessary to have a well defined Gaussian path integral. We  show that  \eq{ybc2} naturally arises when the boundary  path integral over the field configurations defined at  the given return time is carried out. 

It would be interesting to generalize our  results to vector and tensor fields propagating in a classical background. In that case, a function space  with a positive definite inner product  can only be introduced after removing the longitudinal components by gauge fixing. Another important generalization would be to include the scalar metric fluctuations, which is crucial in cosmological perturbation theory (see \cite{pr} for the path integral formulation of cosmological perturbations). There are two main technical obstacles in that case. The first one is the issue of the gauge freedom, which can be handled either by gauge fixing or by working with the gauge invariant variables. The second one is that the exact closed form of the interactions are not known, i.e. one has an infinite series that can be organized in powers of the slow-roll parameters of inflation \cite{mal}. Therefore, first a subset of dominant interactions must be chosen to search for possible non-perturbative effects. Our work is in progress examining these issues.  

\appendix*

\section{The spectral representation and the calculation of the propagators in flat space}

In this Appendix we give some of the details of our computations showing that the functions \eq{f1} and \eq{f2} form an orthonormal set of basis vectors in $L^{(2,2)}(-\infty,0)$ obeying the boundary conditions \eq{rbc1} and \eq{rbc2}. Moreover, we prove that using these eigenfunctions in \eq{gu} gives the standard Green function \eq{fp}. It is instructive and technically much more easier to study $\o=0$ case, where the operator becomes 
\be\label{al}
{\bf L} =\left[\begin{array}{cc}\fr{d^2}{dt^2}&0\\0&-\fr{d^2}{dt^2}\end{array}\right]. 
\ee
The eigenfunctions of $\fr{d^2}{dt^2}$ are  given by $\sin(Et)$, $\cos(Et)$ and $\exp(\pm Et)$. While $\sin(Et)$ and $\cos(Et)$ are allowed in the eigenfunctions of \eq{al} that would yield the delta function normalization, only exponentially decaying function can show up. Since we consider the negative half line $t\in(-\infty,0)$, one should only take the function $\exp(Et)$ with $E\geq 0$. Moreover, once $\sin(Et)$ or $\cos(Et)$ appear in the one entry of the eigenfunction, the other entry must be $\exp(Et)$ due to the minus sign in \eq{al}. 

Taking the above comments  and  the boundary conditions \eq{rbc1} and \eq{rbc2}  into account, it is not too difficult to obtain the following eigenfunctions
\be
u(E,t)=\fr{1}{\sqrt{\pi}}\left[\begin{array}{c}\sin(Et)+\cos(Et)\\ \exp(Et)\end{array}\right],\hs{5} v(E,t)=\fr{1}{\sqrt{\pi}}\left[\begin{array}{c} \exp(Et) \\ \sin(Et)+\cos(Et) \end{array}\right],\label{ae}
\ee
which obey
\be
{\bf L}u(E,t)=-E^2 u(E,t),\hs{5}{\bf L}v(E,t)=E^2 u(E,t),
\ee
where as noted above we take $E\geq 0$. Using
\bea
&&\int_{-\infty}^0\cos(Et)\cos(E't)dt=\fr{\pi}{2}\d(E+E')+\fr{\pi}{2}\d(E-E'),\\
&&\int_{-\infty}^0\sin(Et)\sin(E't)dt=-\fr{\pi}{2}\d(E+E')+\fr{\pi}{2}\d(E-E'),
\eea
and 
\be
\int_{-\infty}^0 \sin(Et)dt=-\fr{1}{E},\label{av1}
\ee
which are valid for any two real numbers $E$ and $E'$, it is a simple exercise to prove that the eigenfunctions are orthonormal
\bea
&&<u(E)|u(E')>=<v(E)|v(E')>=\d(E-E'),\\
&&<u(E)|v(E')>=0,
\eea
and complete
\be
\int_0^\infty dE \left[u(E,t)u^T(E,t')+v(E,t)v^T(E,t')\right]=\d(t-t').
\ee
Recall that the inner product is defined in \eq{rip}. To verify \eq{av1}, which is true in the distributional sense, one may introduce a small convergence factor as
\be
\int_{-\infty}^0 \sin(Et)\,e^{\e\,t}\,dt=-\fr{E}{\e^2+E^2}.
\ee
If this expression is used in a convergent $E$-integral, one may safely take $\e\to0$ limit, which gives \eq{av1}.   

When $\o\not=0$, the general form of the eigenfunctions are given in \eq{sol}. To determine the unknown coefficients in \eq{sol}, one may start applying the boundary conditions \eq{rbc1} and \eq{rbc2}, which leaves only one unknown for $\l>\o^2$ and one for $\l<-\o^2$, and only two unknowns for $\o^2\geq\l\geq-\o^2$.  The normalization
\be
<u(\l)|u(\l')>=\d(\l-\l')
\ee
fixes the eigenfunctions uniquely up to phase factors for $\l>\o^2$ and $\l<-\o^2$ that gives \eq{f1}, while it imposes the following relation 
\be\label{ac}
\left[(\l+\o^2)\sqrt{-\l+\o^2}+(-\l+\o^2)\sqrt{\l+\o^2}\right]a^2+\left[\sqrt{-\l+\o^2}+\sqrt{\l+\o^2}\right]b^2=\fr{1}{\pi},
\ee 
where $a$ and $b$ are two real unknowns that appear in the eigenfunctions for $\o^2\geq\l\geq\o^2$ as
\be
u(\l)=\left[\begin{array}{l} a\sqrt{\l+\o^2}\sin\left(t\sqrt{-\l+\o^2}\right)+b\cos\left(t\sqrt{-\l+\o^2}\right)\\
 a\sqrt{-\l+\o^2}\sin\left(t\sqrt{\l+\o^2}\right)+b\cos\left(t\sqrt{\l+\o^2}\right) 
 \end{array}\right].\label{au}
\ee
In writing \eq{au} we implemented the boundary conditions. Let us remind that $a$ and $b$ are $\l$ dependent parameters. 

To proceed one should check the completeness of the eigenfunctions \eq{comp}. For that it is useful to derive an intermediate integral identity as follows. Let $f(z)$ be the following  complex analytic function
\be\label{fz1}
 f(z)=\fr{\sqrt{z-\o^2}-i\sqrt{z+\o^2}}{2\pi z}\exp\left(t\sqrt{z-\o^2}-it'\sqrt{z+\o^2}\right). 
 \ee
The function has two branch points located at $z=\pm\o^2$. By introducing two cuts as in Fig \ref{fg2}, the square roots become single valued. We define $\sqrt{z\pm\o^2}=\sqrt{r_\pm}\exp(i\th_\pm/2)$, where $r_\pm$ is the distance of $z$ to $\pm\o^2$ and $\th_\pm$ are measured with respect to $x$-axis along clockwise direction. It is easy to see that $f(z)$ is analytic at $z=0$. Consider now the integral of $f(z)$ along the contour in Fig \ref{fg2}. For $t,t'<0$ no contribution comes from the large semicircle. Taking the imaginary part of the integral one may verify the following identity
\bea
&&\int_{\o^2}^\infty\,\fr{d\l}{2\pi\l}e^{t\sqrt{\l-\o^2}}\left[\sqrt{\l-\o^2}\sin(t'\sqrt{\l+\o^2})+\sqrt{\l+\o^2}\cos(t'\sqrt{\l+\o^2})\right]+(t\leftrightarrow t')\nn\\
&&+\int_{-\o^2}^{\o^2}\fr{d\l}{2\pi\l}\left[\sqrt{\l+\o^2}-\sqrt{-\l+\o^2}\right]\times\nn\\
&&\left[\cos(t\sqrt{-\l+\o^2})\cos(t'\sqrt{\l+\o^2})+\sin(t\sqrt{-\l+\o^2})\sin(t'\sqrt{\l+\o^2})\right]=0, \label{aid}
\eea
 which is true for any real $\o$ and $t,t'<0$. 

\begin{figure}
\centerline{\includegraphics[width=8cm]{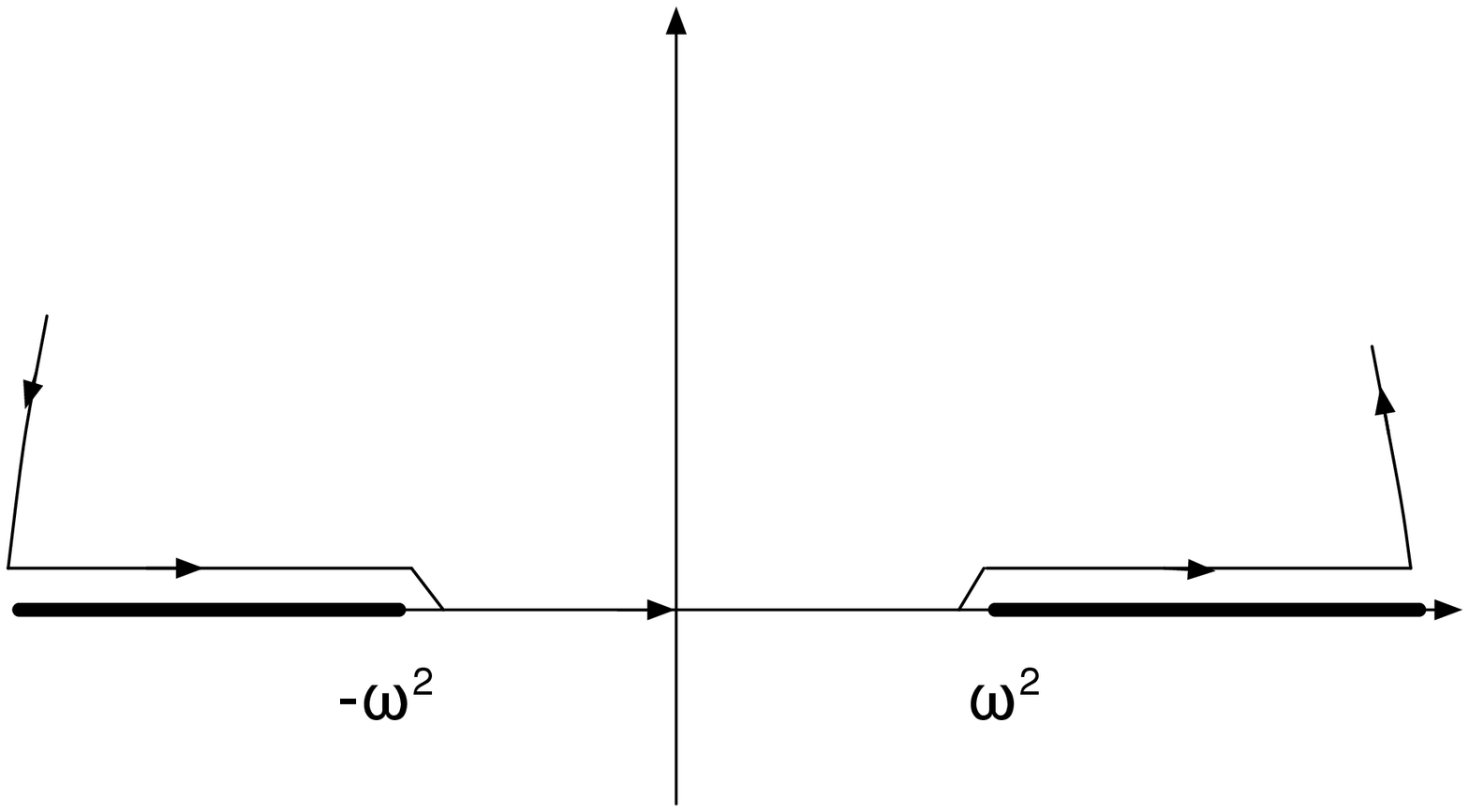}}
\caption{The contour of integration in the complex plane for $f(z)$ in \eq{fz1}.} 
\label{fg2}
\end{figure}

Now, using \eq{f1} and \eq{au} the off-diagonal entry of the completeness relation \eq{comp}  gives
\bea
&&0=\int_{\o^2}^\infty\,\fr{d\l}{2\pi\l}e^{t\sqrt{\l-\o^2}}\left[\sqrt{\l-\o^2}\sin(t'\sqrt{\l+\o^2})+\sqrt{\l+\o^2}\cos(t'\sqrt{\l+\o^2})\right]+(t\leftrightarrow t')\nn\\
&&+\int_{-\o^2}^{\o^2}d\l\left(
\left[b^2\cos(t\sqrt{-\l+\o^2})\cos(t'\sqrt{\l+\o^2})+a^2\sqrt{\o^4-\l^2}\sin(t\sqrt{-\l+\o^2})\sin(t'\sqrt{\l+\o^2})\right]
\right.\nn\\
&&\left.+ab\left[\sqrt{-\l+\o^2}\cos(t\sqrt{-\l+\o^2})\sin(t'\sqrt{\l+\o^2})+\sqrt{\l+\o^2}\sin(t\sqrt{-\l+\o^2})\cos(t'\sqrt{\l+\o^2})\right]\right). \nn
\eea
Comparing this expression with \eq{aid}, we see that the term in the last line containing $ab$ must vanish for the equation to hold. Therefore, there must exist two eigenfunctions having $a$ and $b$ terms, respectively. Using the identity \eq{aid} and the condition \eq{ac}, which is demanded by normalization, it is easy to fix $a$ and $b$ that gives the eigenfunctions in \eq{f2}. 

Since the normalization condition and the off-diagonal component of the completeness relation determine the eigenfunctions uniquely, the diagonal entry of the completeness relation involving  $\d(t-t')$ must be satisfied identically.  The computation can be carried out  straightforwardly although the calculation is cumbersome. Here we would like point out a few crucial steps. It is again useful to derive certain integral identities, which can be used to see cancelation of necessary terms.  Let us define
\be\label{fz2}
f(z)=\fr{\sqrt{z-\o^2}}{2\pi z}\exp\left(-i(t+t')\sqrt{z+\o^2}\right)
\ee
and integrate $f(z)$ in the contour shown in Fig. \ref{fg3}. For $t+t'<0$ the contribution of the large semicircle vanishes. The differences with the previous case are that the pole at $z=0$ has to be avoided and the small circle gives a non-zero contribution even when it shrinks to zero size. Taking the imaginary part of the $f(z)$ integral gives 
\bea
&&\fr{\o}{2}\sin(\o(t+t'))=\label{av2}\\
&&\int_0^{\o^2}\fr{d\l}{2\pi\l}\left[\sqrt{-\l+\o^2}\cos(\sqrt{\l+\o^2}(t+t'))-\sqrt{\l+\o^2}\cos(\sqrt{-\l+\o^2}(t+t'))
\right]\nn\\
&&-\int_{\o^2}^\infty \fr{d\l\sqrt{\l-\o^2}}{2\pi\l}\sin(\sqrt{\l+\o^2}(t+t'))-\int_{\o^2}^\infty \fr{d\l\sqrt{\l+\o^2}}{2\pi\l}\exp(\sqrt{\l-\o^2}(t+t')),\nn
\eea
where the term in the left comes from the shrinking semicircle around $z=0$. One may worry about the convergence of the integral for large $\l$, but it is enough for this identity to hold in the distributional sense. Note that as $\o\to0$, \eq{av2} becomes \eq{av1}. 

\begin{figure}
\centerline{\includegraphics[width=8cm]{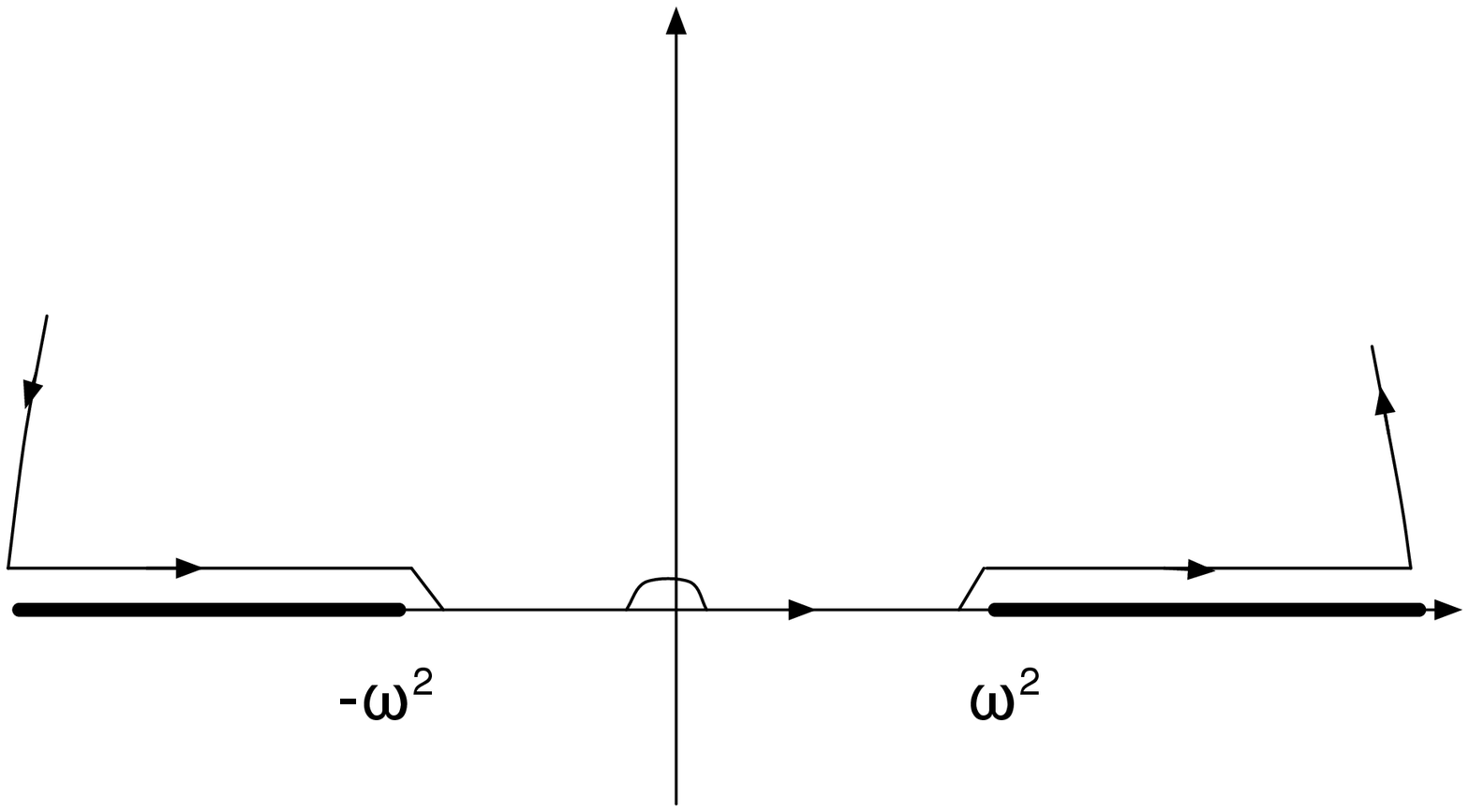}}
\caption{The contour of integration in the complex plane for $f(z)$ in \eq{fz2}.} 
\label{fg3}
\end{figure}

Another integral identity that we use in verifying the diagonal entry of the completeness relation is derived as follows. Define 
\be\label{fz3} 
f(z)=\fr{\o^2}{2\pi z\sqrt{z+\o^2}}\exp\left(-i(t+t')\sqrt{z+\o^2}\right),
\ee
which will be integrated along the contour shown in Fig. \ref{fg4}. Note that there is a single branch point and a single cut. As before, for $t+t'<0$ the large semicircle does not contribute. However, one must calculate the integral along the small semicircle since its contribution is finite as it shrinks down. Taking the real part of that integral gives
\bea
\fr{\o}{2}\sin(\o(t+t'))&&=\int_{0}^\infty\,d\l\,\fr{\o^2}{2\pi\l\sqrt{\l+\o^2}}\cos(\sqrt{\l+\o^2}(t+t'))\nn\\
&&-\int_0^{\o^2}d\l\,\fr{\o^2}{2\pi\l\sqrt{-\l+\o^2}} \cos(\sqrt{-\l+\o^2}(t+t')). \label{av3}
\eea
Using the identities \eq{av2} and \eq{av3}, the diagonal entry of the completeness relation can be verified. 

\begin{figure}
\centerline{\includegraphics[width=8cm]{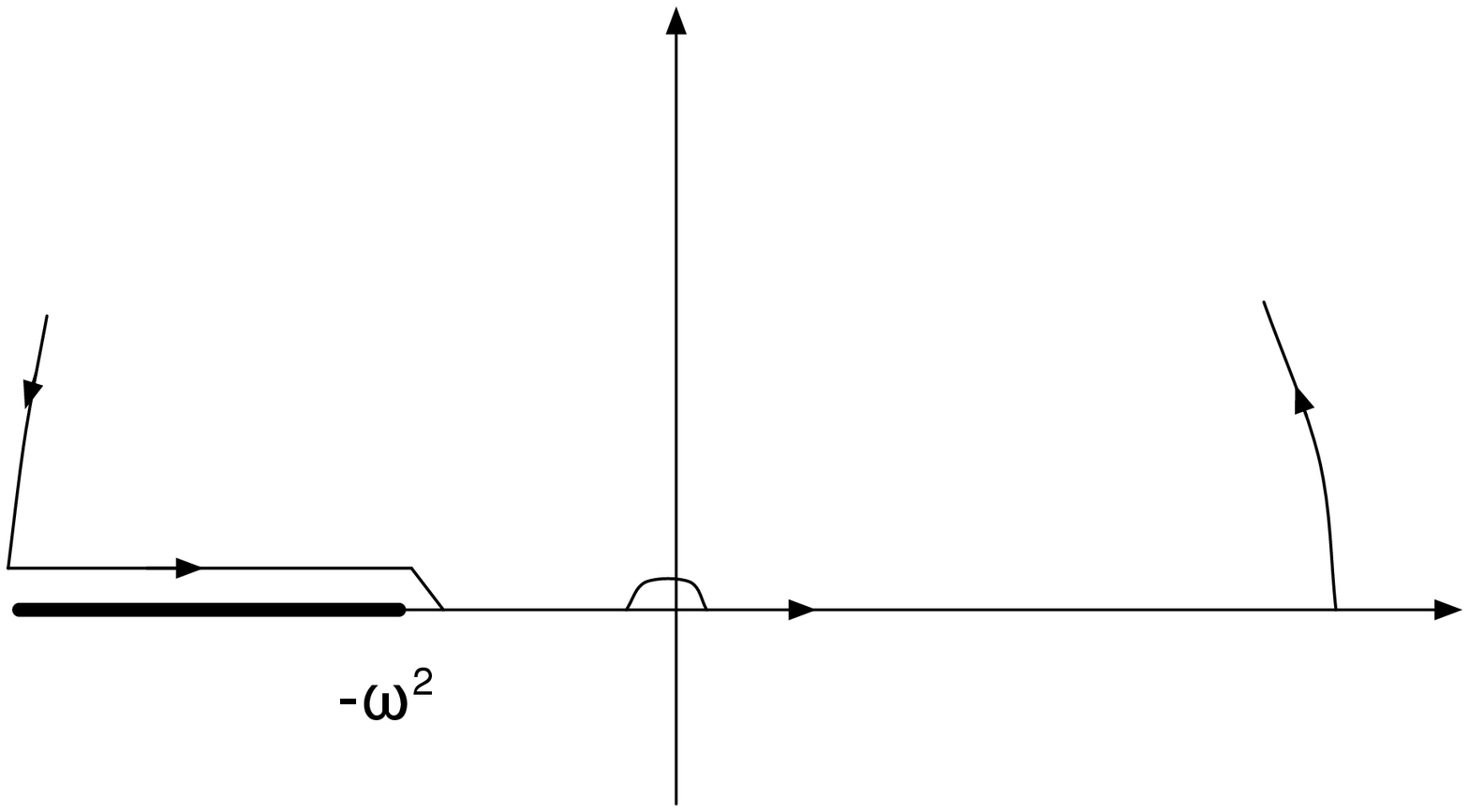}}
\caption{The contour of integration in the complex plane for $f(z)$ in \eq{fz3}.} 
\label{fg4}
\end{figure}

Finally let us present some of the details of the computation that gives \eq{fp} from \eq{gu}. This turns out to be very similar to the verification of the completeness relation \eq{comp};  the crucial difference is the $\l-i\e$ term in the denominator of \eq{gu}. Let us again start from the off-diagonal entry of the Green function. We first observe that for an arbitrary function $f(\l)$ one has 
\bea
\lim_{\e\to0}\int_{-\infty}^{\infty}\,d\l\fr{f(\l)}{\l-i\e}=\int_{-\infty}^{\infty}\,d\l\,f(\l)\,\left[\fr{\l}{\l^2+\e^2}+i\fr{\e}{\l^2+\e^2}\right]=i\pi f(0)+\int_{-\infty}^{\infty}\,d\l\,f(\l)\,\left[\fr{\l}{\l^2+\e^2}\right].\nn\\\label{aie}
\eea
Using this equality for $\Delta^{+-}$ in \eq{gu}, the eigenfunctions \eq{f1} and \eq{f2} give 
\be\label{agim}
{\textrm Im}(\Delta^{+-})=-\fr{i}{2\o}\cos(\o(t-t')).
\ee 
To calculate the real part of $\Delta^{+-}$, one may derive an identity similar to \eq{aid}. Let us define 
\be\label{fz4}
 f(z)=\fr{\sqrt{z-\o^2}-i\sqrt{z+\o^2}}{2\pi z(z-i\e)}\exp\left(t\sqrt{z-\o^2}-it'\sqrt{z+\o^2}\right),
 \ee
and integrate $f(z)$ along the contour shown in Fig. \ref{fg5}. Note that $f(z)$ in \eq{fz4} is equal to \eq{fz1} divided by $z-i\e$ that generates a simple pole at $z=i\e$. Taking the imaginary part of this integral gives the integrand of the real part of $\Delta^{+-}$ multiplied by $\l/(\l^2+\e^2)$, which exactly appears in the identity \eq{aie}. Carefully treating the pole at $z=i\e$ gives
\be
{\textrm Re}(\Delta^{+-})=\fr{1}{2\o}\sin(\o(t-t')).
\ee
Combining with \eq{agim} one then finds
\be\label{agpm}
\Delta^{+-}=-\fr{i}{2\o}\exp(i\o(t-t')).
\ee

\begin{figure}
\centerline{\includegraphics[width=8cm]{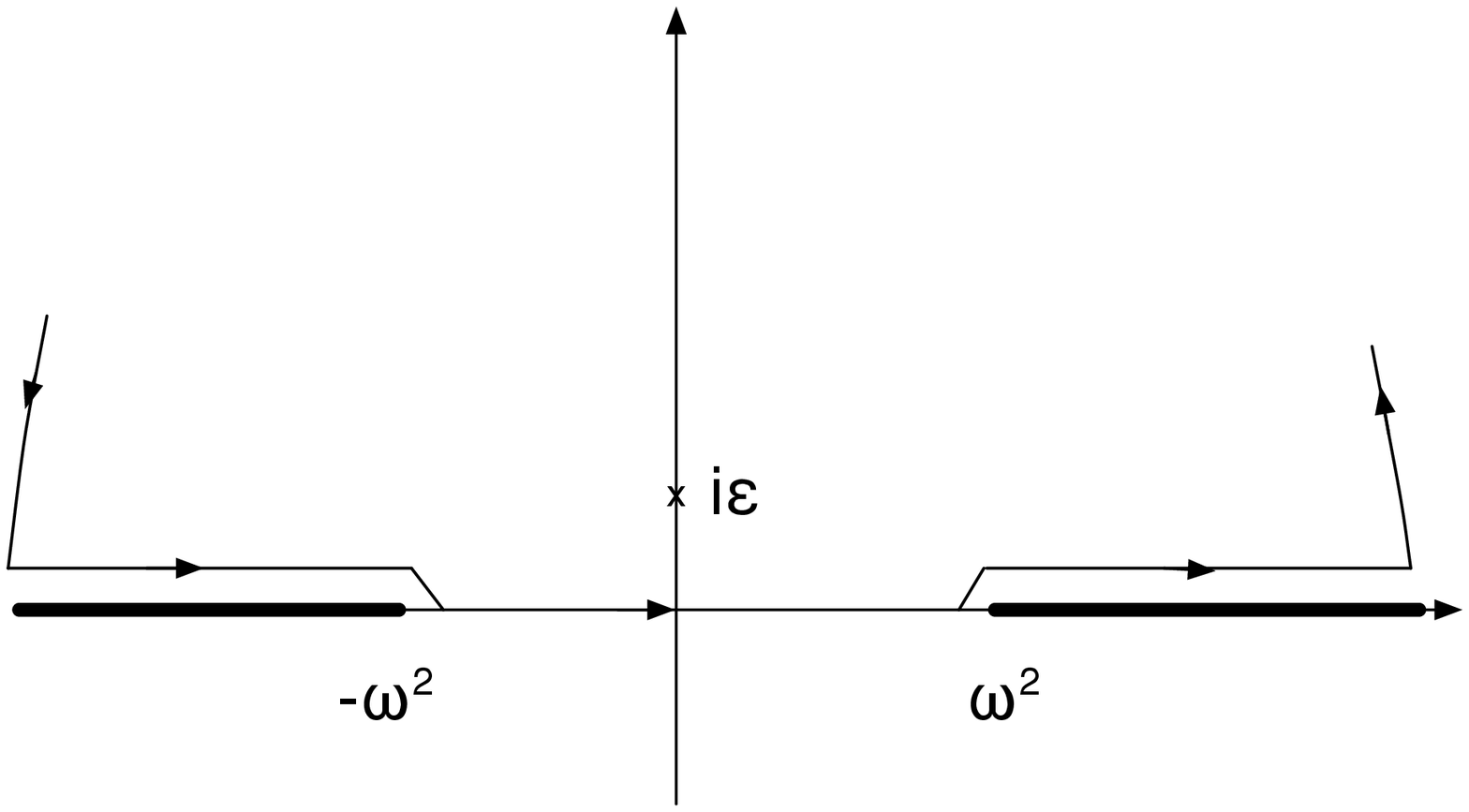}}
\caption{The contour of integration in the complex plane for $f(z)$ in \eq{fz4}.} 
\label{fg5}
\end{figure}

In the calculation of $\Delta^{++}$, one encounters the following integral 
\be\label{aint}
\int_{-\o^2}^\infty\,\fr{d\l}{2\pi\sqrt{\l+\o^2}(\l-i\e)}\cos(\sqrt{\l+\o^2}(t-t')),
\ee
which arises\footnote{At first, the integral limit appears from $\o^2$ to $\infty$. One adds and subtracts to get \eq{aint}.} after using \eq{f1} and \eq{f2} in \eq{gu}. By a change of variable $x=\sqrt{\l+\o^2}$, this integral becomes
\be\label{app}
\int_0^\infty \fr{dx}{\pi(x^2-\o^2-i\e)}\cos(x(t-t'))=\fr{1}{2\pi}\int_{-\infty}^{\infty}\fr{\exp(x(t-t'))}{x^2-\o^2-i\e},
\ee
which then implies
\be\label{agpp}
\Delta^{++}=-i\fr{\th(t-t')}{2\o}e^{i\o(t-t')}-i \fr{\th(t'-t)}{2\o}e^{-i\o(t-t')}+...
\ee
where the dotted term stands for other possible contributions. Let us note that if one drops the $(\l-i\e)$ term in \eq{aint}, it becomes $\d(t-t')$, which is exactly the combination giving the diagonal part of the completeness relation. To see that the remaining terms in \eq{agpp} cancel each other as in the completeness relation, one can modify the identities \eq{av2} and \eq{av3} by multiplying the complex functions \eq{fz2} and \eq{fz3}  by $1/(z-i\e)$. By carefully treating the pole at $z=i\e$ and other contributions, one may verify that all these extra terms cancel each other and \eq{agpp} becomes the final answer. 

The cautious reader would recognize that \eq{fp} is equal to the complex conjugate of the standard in-in propagator. This is not a problem since the propagators in the position space will be the same after the inverse Fourier transformation. In other words, since the correlation functions are real it does not matter if one uses $\Delta$ or $\Delta^*$ as the propagator. Moreover, once $\Delta^{+-}$ and $\Delta^{++}$ are determined, $\Delta^{-+}$ and $\Delta^{--}$ can be fixed from the symmetries of the Green function. 

\begin{acknowledgments}

I would like to thank Stephen Fulling for his help on the essentially self adjoint operators.  I also would like to thank the colleagues in the theoretical high energy physics group at McGill University and especially Robert Brandenberger for their hospitality. This work is supported by T\"{U}B\.{I}TAK B\.{I}DEB-2219 grant. 

\end{acknowledgments}

\end{document}